\documentclass[12pt]{article}

\usepackage{graphicx}
\usepackage[crop=pdfcrop]{pstool}
\usepackage{float}
\usepackage{amsfonts,dsfont}
\usepackage{enumitem}
\usepackage[group-separator={\,}]{siunitx}
\usepackage{mathtools}
\usepackage{chet}

\date{February 2012}

\preprint{UCSD-PTH-12-01}

\title{Scale without Conformal Invariance\\\vspace{6pt} at Three Loops}

\author{Jean-Fran\c{c}ois Fortin, Benjam\'\i{}n Grinstein and Andreas
Stergiou\emails{jffortin@physics.ucsd.edu, bgrinstein@ucsd.edu,
stergiou@physics.ucsd.edu}}

\affiliation{Department of Physics, University of California, San Diego, La
Jolla, CA 92093 USA}

\abstract{We carry out a three-loop computation that establishes the
existence of scale without conformal invariance in dimensional
regularization with the MS scheme in unitary theories in $d=4-\epsilon$
spacetime dimensions.  We also comment on the effects of scheme changes in
theories with many couplings, as well as in theories that live on
non-conformal scale-invariant renormalization group trajectories.
Stability properties of such trajectories are analyzed, revealing both
attractive and repulsive directions in a specific example.  We explain how
our results are in accord with those of Jack \& Osborn on a $c$-theorem in
$d=4$ (and $d=4-\epsilon$) dimensions.  Finally, we point out that limit
cycles with turning points are unlike limit cycles with continuous scale
invariance.}

\newcommand{\CQ}{\ensuremath{\mathcal{Q}}}
\newcommand{\CP}{\ensuremath{\mathcal{P}}}
\DeclareMathOperator{\tra}{tr}

\begin{document}

\maketitle

\begin{center}
  \begin{minipage}{0.8\textwidth}
    \begin{center}
      \textbf{Erratum}
    \end{center}
    \vspace{-10pt}
    The original claim of this paper was that the set of couplings given in
    Eq.~\eqref{ScaleInvPoint} with the result \eqref{qIII} define a theory
    that is scale invariant without being conformal in $D=4-\epsilon$
    spacetime dimensions. This claim is false, as was later realized by the
    authors~\cite{Fortin:2012hn}. The correct interpretation of
    Eq.~\eqref{ScaleInvPoint} is that the theory defined by these couplings
    is fully conformal, although it lives on a limit cycle of the
    traditional dim-reg beta function when the anomalous dimension matrix
    is chosen to be symmetric. This paper remains posted due to this novel
    feature of the solution \eqref{ScaleInvPoint}. For more details the
    reader is referred to~\cite{Fortin:2012hn}.
  \end{minipage}
\end{center}

\newsec{Introduction}
When it was first introduced in its modern form \cite{Polchinski:1987dy},
the question \emph{``Does unitarity and scale invariance imply conformal
invariance?''}\ was mostly of academic interest.  Recent work
\cite{Fortin:2011ks,Fortin:2011sz} showed that scale-invariant theories
display renormalization group (RG) flow recurrent behaviors and have novel
implications for beyond the standard model phenomenology
\cite{Fortin:2011bm}.\foot{For other explorations of scale without
conformal invariance see Refs.~\cite{Hull:1985rc, Iorio:1996ad,
Awad:2000ac, Awad:2000aj, Awad:2000ie, Riva:2005gd, Pons:2009nb,
Dorigoni:2009ra, Jackiw:2011vz, ElShowk:2011gz, Antoniadis:2011gn,
Zheng:2011bp}.} Thus, the existence of scale-invariant theories has deep
consequences, especially with respect to the intuitive understanding of RG
flows as the integrating out of degrees of freedom, and the $c$-theorem.
\emph{``Does unitarity and scale invariance imply conformal invariance?''}\
is therefore not simply a question of academic interest, and to answer it
is of utmost importance.

In Refs.~\cite{Fortin:2011ks,Fortin:2011sz} it was shown that scale does
not necessarily imply conformal invariance in a unitary quantum field
theory (QFT) with enough scalars and fermions at two loops. However, no
completely trustworthy examples have been discovered at this order.  The
failure to find concrete examples at two loops can be understood using the
results of Osborn \cite{Osborn:1989td, Osborn:1991gm} and Jack \& Osborn
\cite{Jack:1990eb}.  In Ref.~\cite{Jack:1990eb} it is argued that, in the
weak-coupling regime, RG flows are gradient flows at two loops.  Hence,
even though scale does not necessarily imply conformal invariance at two
loops, the beta function monomials which could lead to concrete
scale-invariant theories have coefficients that conspire to make all
solutions conformal.  Nothing forbids this from occurring order by order in
perturbation theory.  Therefore, either scale implies conformal
invariance---and the coefficients of the beta function monomials are
tightly constrained, forcing all would-be scale-invariant solutions to be
conformal---or it does not---and recurrent behaviors exist.  Either way,
the answer to the original question leads to important implications
(unexpected structure in the beta functions or the existence of recurrent
behaviors) and the question deserves to be fully investigated.

In this paper we compute the necessary three-loop contributions to the beta
functions to determine if the plausible scale-invariant solutions found in
$d=4-\epsilon$ are eliminated at three loops in the MS scheme, i.e., within
a well-defined renormalization scheme.  Our results show that the
scale-invariant solutions are robust at three loops, and thus open the door
for a $d=4$ scale-invariant example.  Indeed, since scale implies conformal
invariance in pure gauge theories at weak coupling \cite{Polchinski:1987dy,
Jack:1990eb}, the addition of gauge bosons in $d=4$ should not
qualitatively change the $d=4-\epsilon$ results.  For example, the beta
function monomials exhibited below, which lead to an obstruction to the
gradient flow interpretation of the RG flow, are not modified in any way by
the introduction of gauge bosons.  However, to fully answer the question in
$d=4$, one needs the complete three-loop beta functions of theories with
matter and gauge fields, a computation we hope to undertake soon.

It is important to point out that the $c$-theorem discussed in
Refs.~\cite{Osborn:1989td,Osborn:1991gm,Jack:1990eb}, which leads to
$dc/dt=-G_{IJ}\beta^I\beta^J$ with $G_{IJ}$ positive-definite in the weak
coupling regime, is too restrictive.  Indeed, following Osborn
\cite{Osborn:1991gm}, the all-loop proof of the $c$-theorem, which implies
the existence of a monotonically decreasing $c$-function which is constant
only at conformal fixed points, must be modified once spin-one operators of
dimension three are taken into account.  This is exactly the case for
non-conformal scale-invariant theories, since the virial current is such an
operator.  Taking into account the virial current, the analysis is modified
as described in Ref.~\cite[section 3]{Osborn:1991gm}, and leads to
$dc/dt=-(G_{IJ}+\cdots)\beta^IB^J$ where $B^I=\beta^I-\mathcal{Q}^I$ and
$\beta^I=\mathcal{Q}^I$ for non-conformal scale-invariant theories.  Thus,
in its most general form the work of Osborn
\cite{Osborn:1989td,Osborn:1991gm} and Jack \& Osborn \cite{Jack:1990eb}
implies the existence of a $c$-function which is constant at conformal
fixed points ($\beta^I=0$) as well as on scale-invariant trajectories
($B^I=0$).  This is in accord with our three-loop results.

The paper is organized as follows: In section~\ref{expansion}, we discuss
the $\epsilon$ expansion in more detail, showing why the scale-invariant
solutions can be
destabilized at three loops.  We then generate the most general three-loop
beta function for the Yukawa coupling and determine which diagrams
contribute to the virial current.  We finally compute the beta function
coefficients of the relevant diagrams and verify that the virial current
does not vanish at three loops, thus demonstrating the existence of
scale-invariant theories in $d=4-\epsilon$ in a well-defined
renormalization scheme.  Other plausible examples in $d=4-\epsilon$ spacetime
dimensions exhibiting limit cycles are discussed and it
is conjectured that limit cycles and ergodicity are generic in more general
theories.  In section~\ref{scheme}, we
examine scheme changes in theories with many couplings and also on
scale-invariant solutions, showing that, as expected, physical parameters
in $d=4$ do not depend on the renormalization scheme. In
section~\ref{stability} we elucidate the stability properties of
scale-invariant solutions and explicitly verify that the example of
section~\ref{expansion} exhibits both attractive and repulsive
directions.  In section~\ref{c-thm} we return to the arguments of Osborn
\cite{Osborn:1989td,Osborn:1991gm} and Jack \& Osborn \cite{Jack:1990eb}
and show that they are not in contradiction with our results.  Finally, in
section~\ref{trajectories} we contrast our cyclic trajectories with the
trajectories of Ref.~\cite{LeClair:2003hj} which were recently discussed in
connection with the $c$-theorem in Ref.~\cite{Curtright:2011qg} (see also
Ref.~\cite{Morozov:2003ik}).

\newsec{Establishing scale invariance}[expansion]
The results of Refs.~\cite{Fortin:2011ks} were presented in
an expansion in $\epsilon$, similar in spirit to the expansion that reveals
the Wilson--Fisher fixed point.  Let us recall here how that works.  We
consider a model with real scalar fields $\phi_a$ and Weyl spinors $\psi_i$
with quartic scalar self-couplings $\lambda_{abcd}$ and Yukawa couplings
$y_{a|ij}$.  The equations for scale invariance are
\twoseqn{\beta_{abcd}(\lambda,y)&=\CQ_{abcd}\equiv-Q_{a'a}
\lambda_{a'bcd}-Q_{b'b}\lambda_{ab'cd}-Q_{c'c}\lambda_{abc'd}-
Q_{d'd}\lambda_{abcd'}\,,}[ScaleA]{\beta_{a|ij}(\lambda,y)&=
\CP_{a|ij}\equiv-Q_{a'a}y_{a'|ij}-P_{i'i}y_{a|i'j}
-P_{j'j}y_{a|ij'}\,,}[ScaleB][Scale]
where $\beta_{abcd}=-d\lambda_{abcd}/dt$ and $\beta_{a|ij}=-dy_{a|ij}/dt$
are the beta functions for the coupling constants,\foot{With our
conventions RG time increases as we flow to the IR, $t=\ln(\mu_0/\mu)$.}
$Q_{ab}$ is antisymmetric and $P_{ij}$ anti-Hermitian. To proceed, we solve
Eqs.~\Scale for the coefficients of $\lambda,y,Q$ and $P$ in an $\epsilon$
expansion,
\eqn{\lambda_{abcd}=\sum_{n\geq1}\lambda_{abcd}^{(n)}\epsilon^n,\qquad
y_{a|ij}=\sum_{n\geq1}y_{a|ij}^{(n)}\epsilon^{n-\frac{1}{2}},\qquad
Q_{ab}=\sum_{n\geq2}Q_{ab}^{(n)}\epsilon^n,\qquad
P_{ij}=\sum_{n\geq2}P_{ij}^{(n)}\epsilon^n\,.}
Scale-invariant solutions are solutions of Eqs.~\Scale with non-vanishing
$Q$ and/or $P$.

\subsec{Limit Cycle in \texorpdfstring{$d=4-\epsilon$}{d=4-epsilon}: Model
with \texorpdfstring{$2$}{2} Scalars and \texorpdfstring{$2$}{2} Fermions}
For the remainder of this section we will work with a theory of two real
scalars and two Weyl fermions, canonical kinetic terms and interactions
described by
\begin{multline}\label{ttpotential}
V=\tfrac{1}{24}\lambda_1\phi_1^4+\tfrac{1}{24}\lambda_2\phi_2^4
+\tfrac{1}{4}\lambda_3\phi_1^2\phi_2^2
+\tfrac{1}{6}\lambda_4\phi_1^3\phi_2+
\tfrac{1}{6}\lambda_5\phi_1\phi_2^3+(\tfrac{1}{2}y_1\phi_1\psi_1\psi_1\\
+\tfrac{1}{2}y_2\phi_2\psi_1\psi_1+\tfrac{1}{2}y_3\phi_1\psi_2\psi_2
+\tfrac{1}{2}y_4\phi_2\psi_2\psi_2
+y_5\phi_1\psi_1\psi_2+y_6\phi_2\psi_1\psi_2+\text{h.c.})\,.
\end{multline}
This is the simplest weakly-coupled unitary example in $d=4-\epsilon$ with
a well-behaved bounded-from-below scalar potential. For this model $Q$ is
$2\times2$ antisymmetric and $P$ $2\times2$ anti-Hermitian:
\eqn{Q=\begin{pmatrix}0&q\\-q&0\end{pmatrix}\qquad\text{and}\qquad
P=\begin{pmatrix}ip_1&p_3+ip_4\\-p_3+ip_4&ip_2\end{pmatrix},}
where $q$ and $p_{1,\ldots,4}$ are real.

\subsubsec{The two-loop computation}
To start our computation we solve Eq.~\ScaleB at order $\epsilon^{3/2}$.
The result is used in Eq.~\ScaleA which is then solved at order
$\epsilon^2$.  This is a system of coupled \emph{nonlinear} equations and,
as such, it has many solutions $y_{a|ij}^{(1)}$ and $\lambda_{abcd}^{(1)}$,
some of them consistent with unitarity and boundedness of the scalar
potential, while others not.  Additionally, some of these solutions lead to
conformal fixed points, while others allow for nonzero $q$, at least in
principle.

\begin{figure}[H]
  \centering
  \pstool{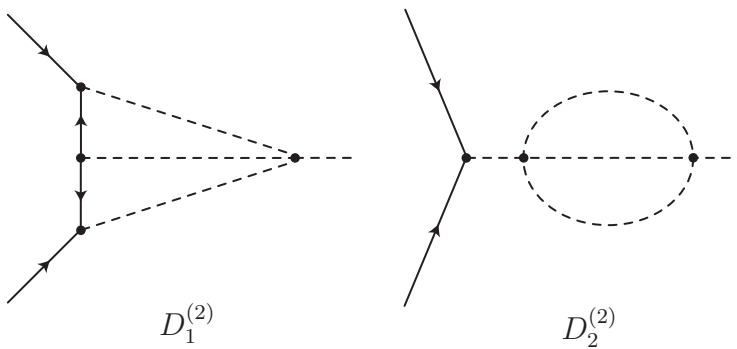}{\psfrag{a}[][]{$D^{(2)}_1$}
                                \psfrag{b}[][]{$D^{(2)}_2$}}
  \caption{Diagrams that contribute to $q$ at two-loop
  order.}
  \label{TwoLDiags}
\end{figure}

At two-loop order solutions $y_{a|ij}^{(1)}$ and $\lambda_{abcd}^{(1)}$ of
the previous order are used to solve Eq.~\ScaleB at order $\epsilon^{5/2}$,
and Eq.~\ScaleA at order $\epsilon^3$.  This is now a system of coupled
\emph{linear} equations,\foot{For all higher orders in $\epsilon$ one only
gets systems of coupled linear equations.} from which the unknowns
$y_{a|ij}^{(2)}$ and $\lambda_{abcd}^{(2)}$ are determined.  For most
$y_{a|ij}^{(1)}$ and $\lambda_{abcd}^{(1)}$ the unknown $q^{(2)}$ is equal
to zero, but for certain $y_{a|ij}^{(1)}$ and $\lambda_{abcd}^{(1)}$, i.e.,
for possible scale-invariant solutions, it is found to be equal to a linear
combination of coefficients of monomials in $\beta_{a|ij}$.  More
specifically, the diagrams that contribute to $q$ at two loops are shown in
Fig.~\ref{TwoLDiags}.

\noindent Let $b_1$ and $b_2$ be their coefficients in the beta function
for the Yukawa coupling,
\eqn{(16\pi^2)^2\beta^{(\text{2-loop})}_{a|ij}\supset
b_1y^{\phantom{*}}_{b|ik}y^*_{c|k\ell}y^{\phantom{*}}_{d|\ell j}
\lambda_{abcd}+b_2y_{b|ij}\lambda_{bcde}\lambda_{acde}.}
Then (we omit the prefactor here since it is not relevant for the
discussion),
\eqn{q^{(2)}\propto b_1+24b_2,}[TwoLoopq]
and because $b_1=-2$ and $b_2=\frac{1}{12}$ we find $q^{(2)}=0$. We also
find that $p_4^{(2)}$ is undetermined, while $p_{1,2,3}^{(2)}=0$. The
freedom in the fermion part of the virial current is related to the
enhanced symmetry
\eqn{
\begin{pmatrix}\psi_1\\\psi_2\end{pmatrix}\rightarrow\begin{pmatrix}
\cos\theta&i\sin\theta\\i\sin\theta&\cos\theta\end{pmatrix}
\begin{pmatrix}\psi_1\\\psi_2\end{pmatrix}.}

As we already mentioned, the failure to find trustworthy non-conformal
scale-invariant solutions at two loops can be explained by the gradient
flow property of the RG flow at weak coupling described in
Ref.~\cite{Jack:1990eb}.  Note that here, contrary to the case of conformal
fixed points, $q^{(3)}\neq0$ at two-loop order. However, the three-loop
contributions to the beta functions can very well conspire to set
$q^{(3)}=0$, and thus restore conformal invariance. (As we will demonstrate
in the next subsection, this does not happen. The fact that $q^{(2)}=0$ is
merely an accident.)

An interesting observation at this point is that if $q^{(2)}=0$ were not an
accident, then that would directly imply that the conformal symmetry
somehow relates vertex corrections and wavefunction renormalizations. This
is obvious from the fact that the first diagram in Fig.~\ref{TwoLDiags}
contributes to the residue of the $1/\epsilon$ pole of $Z_y$, while the
second to the residue of the $1/\epsilon$ pole of $Z_\phi$. This would be
reminiscent, e.g., of the Ward identity for charge conservation in QED.

A point on the candidate scale-invariant trajectory is given by
\eqna{\lambda_1&=\tfrac{8(\num{7087} +
\num{357}\sqrt{\num{52953}})}{\num{102885}}\pi^2\epsilon
+\tfrac{2(\num{490537743519}+\num{468277825}\sqrt{\num{52953}})}
{\num{408605205375}}\pi^2\epsilon^2+\cdots\,,\\
\lambda_2&=\tfrac{64(\num{6346}+9\sqrt{\num{52953}})}{\num{102885}}
\pi^2\epsilon+\tfrac{17(\num{11340943081}+\num{57223077}\sqrt{\num{52953}})}
{\num{136201735125}}\pi^2\epsilon^2+\cdots\,,\\
\lambda_3&=-\tfrac{\num{272}(\sqrt{\num{52953}}-57)}{\num{102885}}
\pi^2\epsilon+\tfrac{\num{291302437755}-\num{3043364867}\sqrt{\num{52953}}}
{\num{817210410750}}\pi^2\epsilon^2+\cdots\,,\\
\lambda_4&=\tfrac{32\sqrt{323(\num{757}-3\sqrt{\num{52953}})}}
{\num{102885}}\pi^2\epsilon+\tfrac{13\sqrt{\splitfrac{
\scriptstyle{\num{190447787}(\num{13924269796644128925781}}}{\scriptstyle{-\num{49509459494439826531}}\sqrt{\num{52953}})}}}
{\num{55843528611660750}}\pi^2\epsilon^2+\cdots\,,\\
\lambda_5&=\tfrac{\num{272}\sqrt{323(\num{757}
-3\sqrt{\num{52953}})}}{\num{102885}}\pi^2\epsilon
+\tfrac{\sqrt{\splitfrac{\scriptstyle{\num{571343361}}
(\num{652474762867234518381407}}{\scriptstyle{-\num{663663219013252691017
}\sqrt{\num{52953}})}}}}
{\num{19709480686468500}}\pi^2\epsilon^2+\cdots\,,\\
y_1&=-y_3=\tfrac{2\sqrt{10}}{5}\pi\epsilon^{1/2}+\tfrac{\sqrt{10}(\num{175503}+\num{442}\sqrt{\num{52953}})}{\num{3249000}}\pi\epsilon^{3/2}+\cdots\,,
}[ScaleInvPoint]
where the remaining couplings vanish at this point.

One can check that Eqs.~\Scale are satisfied on this scale-invariant
trajectory with the help of the two-loop beta functions of
Ref.~\cite{Jack:1984vj}.  Since $q=\mathcal{O}(\epsilon^3)$ we need the
three-loop Yukawa beta functions in order to establish that this solution
is indeed a scale-invariant trajectory in dimensional regularization.

\begin{figure}[H]
  \centering
  \pstool{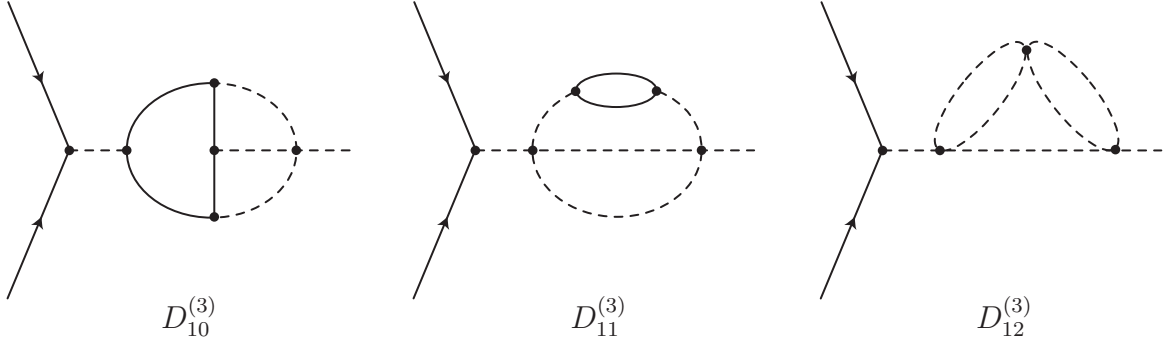}{
          \psfrag{a}[][]{$D^{(3)}_{10}$}
          \psfrag{b}[][]{$D^{(3)}_{11}$}
          \psfrag{c}[][]{$D^{(3)}_{12}$}}
  \caption{Diagrams that contribute to $q$ at three-loop order.}
  \label{ThreeLDiags}
\end{figure}

\subsubsec{The three-loop computation}
There is a large number of diagrams that contribute to $\beta_{a|ij}$ at
three loops. (We use the \emph{Mathematica} package
\href{http://www.feynarts.de/}{\texttt{FeynArts}} to automatically generate
all required diagrams.)  Each diagram corresponds to a unique combination
of coupling constants, and an examination of all of them reveals that those
that contribute to $q^{(3)}$ are the ones shown in Fig.~\ref{ThreeLDiags}.
\begin{figure}[H]
  \centering
  \pstool{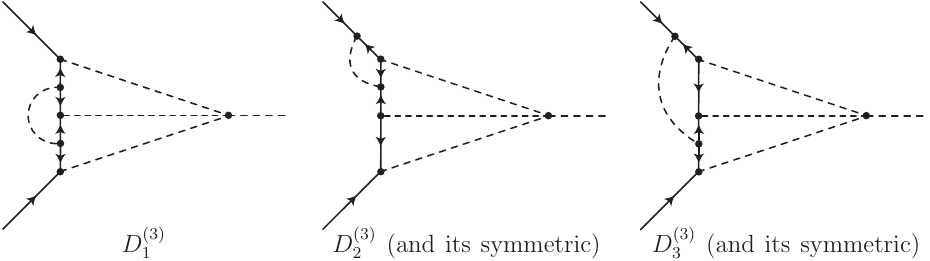}{
          \psfrag{a}[][]{$D^{(3)}_1$}
          \psfrag{b}[][]{$D^{(3)}_2$ (and its symmetric)}
          \psfrag{c}[][]{$D^{(3)}_3$ (and its symmetric)}}
\end{figure}
\begin{figure}[H]
  \centering
  \pstool{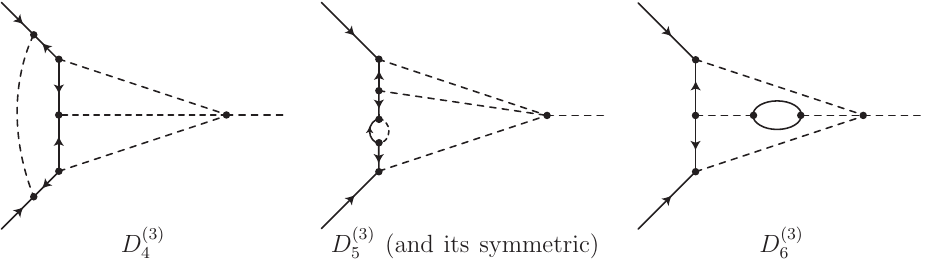}{
          \psfrag{a}[][]{$D^{(3)}_4$}
          \psfrag{b}[][]{$D^{(3)}_5$ (and its symmetric)}
          \psfrag{c}[][]{$D^{(3)}_6$}}
\end{figure}
\begin{figure}[H]
  \centering
  \pstool{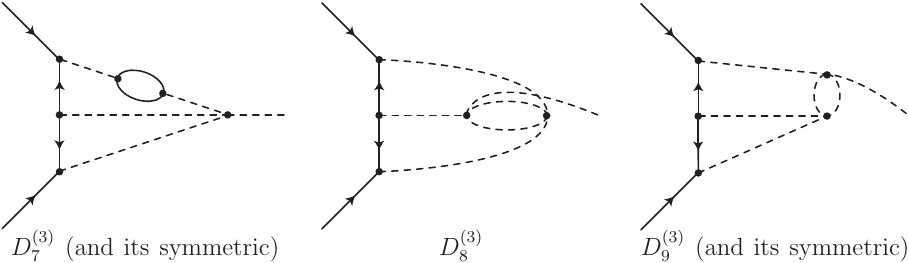}{
          \psfrag{a}[][]{$D^{(3)}_7$ (and its symmetric)}
          \psfrag{b}[][]{$D^{(3)}_8$}
          \psfrag{c}[][]{$D^{(3)}_9$ (and its symmetric)}}
\end{figure}

Note that the diagrams in Fig.~\ref{ThreeLDiags} are specific to the
example of this subsection.  More complicated models might involve more
diagrams.  However, we have checked that precisely the same diagrams
contribute to $q^{(3)}$ in the model with two scalars and one Weyl spinor
of Ref.~\cite{Fortin:2011ks}. It is interesting to point out that very few
of the $\sim\! 200$ diagrams in $\beta_{a|ij}$ contribute to $q^{(3)}$, and
that the ones that do involve both Yukawa and quartic vertices.  The same
situation is encountered at two loops, and we conjecture that it holds to
all orders in perturbation theory.  This is also motivated by comments in
Ref.~\cite{Wallace:1974dy} regarding the ``interference'' between
successive loop orders in the calculation of a potential for a gradient
flow (see also section~\ref{c-thm} below).  It is also curious that the
diagrams of Fig.~\ref{ThreeLDiags} have an (obvious) topological relation
to the diagrams of Fig.~\ref{TwoLDiags}.

The diagrams of Fig.~\ref{ThreeLDiags} have simple poles in $\epsilon$ and
so they contribute to the Yukawa beta function at three loops:
\eqn{(16\pi^2)^3\beta^{\text{(3-loop)}}_{a|ij}\supset
c_1 y^{\phantom{*}}_{b|ik}y^*_{c|k\ell}y^{\phantom{*}}_{d|\ell m}y^*_{c|mn}
y^{\phantom{*}}_{e|nj}\lambda_{abde}+\cdots+c_{12}y_{b|ij}
\lambda_{bcde}\lambda_{cdf\!g}\lambda_{ae\!f\!g}.}
The three-loop analog of Eq.~\TwoLoopq is then (again omitting the
prefactor)\foot{For the model of two scalars and one Weyl spinor of
Ref.~\cite{Fortin:2011ks} the expression for $q^{(3)}$ is
\vspace{-6pt}
\eqn{q^{(3)}\propto -219 + 12(4c_1+2c_2+2c_3+c_4+2c_5+2c_6+4c_7) +
4(5c_8+10c_9+6c_{10}+10c_{11}+187c_{12}).}
\vspace{-12pt}
}
\eqn{q^{(3)}\propto
-71+3(c_1+2c_2+2c_3+c_4+2c_5+4c_6+8c_7)+
  4(c_8+2c_9+3c_{10}+4c_{11}+58c_{12}),}[ThreeLoopq]
where the constant piece comes from contributions to $q^{(3)}$ from the
previous order.

To compute these three-loop diagrams we implemented the algorithm of
Ref.~\cite{Chetyrkin:1997fm}.\foot{We would like to thank M.\ Misiak for
pointing us to this reference.}  There, IR divergences are regulated by
introducing a spurious mass parameter through an exact decomposition of the
massless propagator, and the calculation proceeds with properly choosing a
loop momentum, regarding it as large, and expanding with respect to it the
remaining two-loop subintegral, for which the chosen momentum is external.
Remarkably, the authors of Ref.~\cite{Chetyrkin:1997fm} manage to construct
explicit formulas for the pole parts of all three-loop scalar integrals.
The implementation of their algorithm is straightforward, e.g., in
\emph{Mathematica}, but one must be very careful to take into account all
required counterterms, including the ones introduced by the IR regulator.
To test our implementation, we verified the two-loop result of
Ref.~\cite{Jack:1984vj} for $\beta_{a|ij}$, and also part of the three-loop
result for the beta function of the quartic coupling in a multi-flavor
theory of scalars found in Ref.~\cite{Jack:1990eb}.  We also performed
explicit computations of a couple of diagrams.

From the diagrams of Fig.~\ref{ThreeLDiags} we find
\eqna{c_1&=3, & c_2&=-1, & c_3&=2, & c_4&=5, & c_5&=\tfrac12, &
c_6&=\tfrac32,\\ c_7&=\tfrac12, & c_8&=\tfrac32, & c_9&=\tfrac12, &
c_{10}&=\tfrac58, & c_{11}&=-\tfrac{5}{32}, & c_{12}&=-\tfrac{1}{16}.}
Restoring the prefactor, then, Eq.~\ThreeLoopq gives\foot{For the model of
two scalars and one Weyl spinor of Ref.~\cite{Fortin:2011ks} we find
\eqn{q^{(3)} =\frac{35 \sqrt{\num{34706} (\num{3601}+6\sqrt{\num{419802}})}}
{\num{2489696256}}\approx 2\times 10^{-4} .}[qIII]
}
\eqn{q^{(3)}=\frac{\sqrt{\num{323} (\num{757}-3
\sqrt{\num{52953}})}}{\num{2057700}}\approx 7\times 10^{-5}.}
Since $q^{(3)}\neq 0$ we have established the
existence of theories that are scale but not conformally invariant!
We expect that theories in $d=4$ can also display scale
without conformal invariance.

To summarize, it is important to emphasize that the distinction between
scale-invariant and conformal solutions of Eqs.~\Scale at the two-loop
level is that, for the latter, $q^{(\geq3)}=0$ already at two loops. Higher
loops are expected to slightly modify the critical values of the couplings,
while preserving $q=0$.  But there are solutions for which $q^{(\geq
3)}\neq 0$ already at two loops.  As a result, the nature of these
solutions is uncertain, and a higher-loop calculation is needed.  Even
without that calculation, though, it should be clear that not all solutions
to Eqs.~\Scale can be declared conformal with the same confidence and the
three-loop computation we present here shows that indeed non-conformal
scale-invariant solutions exist.

\begin{figure}[ht]
  \centering
  \pstool{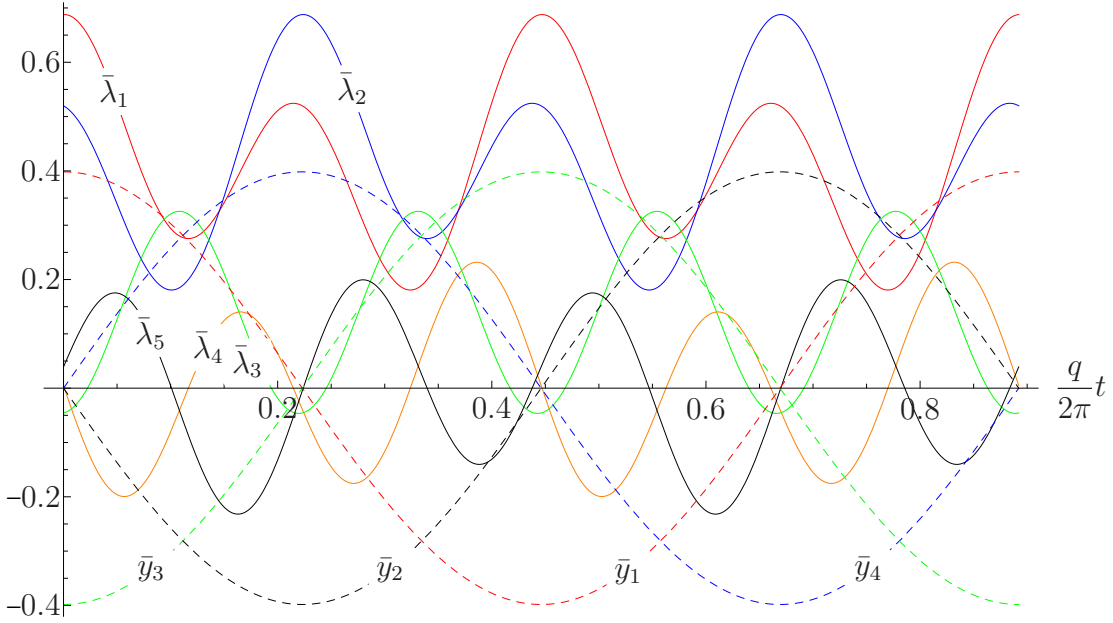}{\psfrag{t}{$\dfrac{q}{2\pi}t$}
  \psfrag{0.2}[][]{$0.2$} \psfrag{0.4}[][]{$0.4$} \psfrag{0.6}[][]{$0.6$}
  \psfrag{2}[][]{0.2} \psfrag{4}[][]{0.4} \psfrag{6}[][]{0.6}
  \psfrag{8}[][]{0.8} \psfrag{l}[][]{\colorbox{white}{$\bar{\lambda}_1$}}
  \psfrag{m}[][]{\colorbox{white}{$\bar{\lambda}_2$}}
  \psfrag{n}[][]{\colorbox{white}{$\bar{\lambda}_3$}}
  \psfrag{o}[][]{\colorbox{white}{$\bar{\lambda}_4$}}
  \psfrag{p}[][]{\colorbox{white}{$\bar{\lambda}_5$}}
  \psfrag{q}[][]{\colorbox{white}{$\bar{y}_1$}}
  \psfrag{r}[][]{\colorbox{white}{$\bar{y}_4$}}
  \psfrag{s}[][]{\colorbox{white}{$\bar{y}_3$}}
  \psfrag{u}[][]{\colorbox{white}{$\bar{y}_2$}} }
  \caption{RG evolution of the couplings in the model with two real scalars
and two Weyl fermions on a scale-invariant limit cycle as a function of RG
time.  Here $\epsilon=0.01$.}\label{Fig:CoupPlot2s2f}
\end{figure}

Since there is only one oscillation frequency the scale-invariant
trajectory is a limit cycle.  The RG evolution of the couplings along the
limit cycle is easily determined from Eqs.~\Scale and is shown in
Fig.~\ref{Fig:CoupPlot2s2f} for $\epsilon=0.01$.
Notice that all phases can be rotated away and thus the model does not
violate CP.  Moreover the minimum of the scalar potential is located at the
origin of field space.  As expected, these statements (boundedness of the
scalar potential, CP conservation, location of the vacuum in field space)
are invariant along the limit cycle.

\newsec{Other plausible examples}

\subsec{Limit Cycle in \texorpdfstring{$d=4-\epsilon$}{d=4-epsilon}: Model
with \texorpdfstring{$3$}{3} Scalars and \texorpdfstring{$2$}{2} Fermions}

The next simplest example in $d=4-\epsilon$ with a scalar potential which
is bounded from below is described by a theory of three real scalars and
two Weyl fermions, with canonical kinetic terms and interactions described
by
\begin{multline*}
V=\tfrac{1}{24}\lambda_1\phi_1^4+\tfrac{1}{24}\lambda_2\phi_2^4
+\tfrac{1}{24}\lambda_3\phi_3^4+\tfrac{1}{4}\lambda_4\phi_1^2\phi_2^2
+\tfrac{1}{4}\lambda_5\phi_1^2\phi_3^2
+\tfrac{1}{4}\lambda_6\phi_2^2\phi_3^2
+\tfrac{1}{6}\lambda_7\phi_1^3\phi_2
+\tfrac{1}{6}\lambda_8\phi_1^3\phi_3
+\tfrac{1}{6}\lambda_9\phi_1\phi_2^3\\
+\tfrac{1}{6}\lambda_{10}\phi_2^3\phi_3
+\tfrac{1}{6}\lambda_{11}\phi_1\phi_3^3
+\tfrac{1}{6}\lambda_{12}\phi_2\phi_3^3
+\tfrac{1}{2}\lambda_{13}\phi_1^2\phi_2\phi_3
+\tfrac{1}{2}\lambda_{14}\phi_1\phi_2^2\phi_3
+\tfrac{1}{2}\lambda_{15}\phi_1\phi_2\phi_3^2\\
+(\tfrac{1}{2}y_1\phi_1\psi_1\psi_1+\tfrac{1}{2}y_2\phi_2\psi_1\psi_1
+\tfrac{1}{2}y_3\phi_3\psi_1\psi_1+\tfrac{1}{2}y_4\phi_1\psi_2\psi_2
+\tfrac{1}{2}y_5\phi_2\psi_2\psi_2\\+\tfrac{1}{2}y_6\phi_3\psi_2\psi_2
+y_7\phi_1\psi_1\psi_2+y_8\phi_2\psi_1\psi_2
+y_9\phi_3\psi_1\psi_2+\text{h.c.})\,.
\end{multline*}
Here the unknown parameters $Q_{ab}$ and $P_{ij}$ in the virial current are
given by
\eqn{
Q=\begin{pmatrix}0&q_1&q_2\\-q_1&0&q_3\\-q_2&-q_3&0\end{pmatrix},\qquad
P=\begin{pmatrix}ip_1&p_3+ip_4\\-p_3+ip_4&ip_2\end{pmatrix},}
where $q_{i=1,\ldots,3}$ and $p_{i=1,\ldots,4}$ are real.  All the scalar
quartic couplings, $\lambda_{1,\ldots,15}$, and two of the Yukawa
couplings, $y_1$ and $y_4$, do not vanish on the scale-invariant
trajectory.  Due to its lengthy form we do not give here the explicit
$\epsilon$-expansion.  Its exact knowledge does not lead to a better
understanding of the physics and, moreover, the $\epsilon$-expansion can
easily be determined from the two-loop beta functions of
Ref.~\cite{Jack:1984vj} and our new three-loop results.
The non-vanishing virial current parameters on
this scale-invariant trajectory are $q_1$, $q_2$ and $p_4$.  The
$\epsilon$-expansion for $q_1$ and $q_2$ are distinct while, again, $p_4$
is undetermined and corresponds to the enhanced symmetry
\eqn{
\begin{pmatrix}\psi_1\\\psi_2\end{pmatrix}\rightarrow
  \begin{pmatrix}\cos\theta&i\sin\theta\\
    i\sin\theta&\cos\theta
  \end{pmatrix}
  \begin{pmatrix}\psi_1\\
    \psi_2
  \end{pmatrix}
  }
of the scale-invariant trajectory.

Since on this scale-invariant trajectory the oscillation frequencies are
$\pm\sqrt{q_1^2+q_2^2+q_3^2}$ and $0$, the scale-invariant trajectory is
also a limit cycle.  Again, the model has a bounded-from-below scalar
potential, does not violate CP and has a minimum at the origin of field
space.

\subsec{Limit Cycle and Ergodicity in
\texorpdfstring{$d=4-\epsilon$}{d=4-epsilon}: Model with
\texorpdfstring{$N_S>3$}{Ns>3} Scalars and \texorpdfstring{$N_F>2$}{Nf>2}
Fermions}

Up to now the models in $d=4-\epsilon$ spacetime dimensions display
scale-invariant trajectories that are limit cycles.  Although the virial
current has enough freedom to lead to several oscillation frequencies, in
both models the non-trivial part of $P_{ij}$ vanishes and thus the
oscillation frequencies are solely obtained from $Q_{ab}$.  For two and
three real scalars it is thus impossible to get scale-invariant
trajectories that exhibit ergodicity.  Indeed, the eigenvalues of $Q_{ab}$
are $\{\pm iq_1\}$ and $\{0,\pm i\sqrt{q_1^2+q_2^2+q_3^2}\}$ for two and
three real scalars respectively, implying limit cycles.  Eigenvalues of
antisymmetric matrices with real entries always come in pairs $\pm
i\omega$, except in the case where the dimensionality of the matrix is odd,
where, in addition, there is a zero eigenvalue.  Therefore, assuming
$P_{ij}=0$, four or more real scalars are necessary to obtain ergodic
behaviors.  For example, since the $\epsilon$-expansion for the $q_i$ are
generically distinct it is expected that the model with four real scalars
and two Weyl fermions will display both limit cycles and ergodic behavior
as a function of $\epsilon$.

We therefore conjecture that ergodic behavior in $d=4-\epsilon$ spacetime
dimensions occurs in models with $N_S>3$ real scalars and $N_F>1$ Weyl
fermions.  Unfortunately, due to the large number of couplings (for example
the model with four real scalars and two Weyl fermions has
$\frac{N_S(N_S+1)(N_S+2)(N_S+3)}{4!}+2\times N_S\frac{N_F(N_F+1)}{2}=59$
real couplings) the computing time necessary to generate the three-loop beta
functions becomes excessive and we have not pursued this direction further.

\newsec{Renormalization-scheme changes}[scheme]

\subsec{Scheme changes and conformal fixed points: the one-coupling case}
Let us first review the effects of scheme changes in conformal theories.
The simple case of a theory with only one coupling has been investigated
long ago in Ref.~\cite{Gross:1975vu}.  Under a scheme change, the coupling
$g$ and the wavefunction renormalization $Z(g)$ become
\eqna{g&\to \tilde{g}(g)=g+\mathcal{O}(g^3),\\
Z^{1/2}(g)&\to \widetilde{Z}^{1/2}(\tilde{g})=
Z^{1/2}(g)F(g),}[Schemechangeone]
where $F(g)=1+\mathcal{O}(g^2)$ and $F\neq0$ for all $g$.  In the new
scheme $\tilde{g}$ is equal to $g$ at lowest order since the coupling is
unambiguous at the classical level.  The same is true for the wavefunction
renormalization as well.  Therefore, since\foot{We use
$\phi_\text{B}=Z^{1/2}(g)\phi_\text{R}$.}
\eqna{\beta(g)&=-\frac{dg}{dt},\\
\gamma(g)&=-Z^{-1/2}(g)\frac{d Z^{1/2}(g)}{dt},}
the new beta function and anomalous dimension are related to the old beta
function and anomalous dimension through
\eqna{\tilde{\beta}(\tilde{g})&=
  \beta(g)\frac{\partial\tilde{g}}{\partial g},\\
\tilde{\gamma}(\tilde{g})&=
  \gamma(g)+F^{-1}(g)\beta(g)\frac{\partial F(g)}{\partial g}.}[RGnewoldone]
Although the RG functions depend strongly on the renormalization scheme,
properties that have physical consequences must be independent of the
scheme.  Such properties are:
\begin{enumerate}[label=(\Roman{enumi}$'$)]
    \renewcommand{\labelenumi}{(\Roman{enumi})}
  \item\label{prop1} The existence of a conformal fixed point;
  \item\label{prop2} The anomalous dimension at a conformal fixed point,
  which determines the scaling behavior of Green functions;
  \item\label{prop3} The first derivative of the beta function at a
  conformal fixed point, which determines the sign\foot{Note that the sign
  determines the character (attractive or repulsive) of the conformal fixed
  point.} and rate of approach of the coupling to the conformal fixed point
  and thus modifies asymptotic formulae;
  \item\label{prop4} The first two coefficients in the beta function, which
  govern the UV or IR asymptotics of the coupling;
  \item\label{prop5} The first coefficient in the anomalous dimension,
  which controls the scale factor of the field in the far UV or IR.
\end{enumerate}
These properties all follow from Eqs.~\RGnewoldone and the form of
$\tilde{g}(g)$ and $F(g)$.

\subsec{Scheme changes and conformal fixed points: the multi-coupling case}
When the theory has more than one coupling, a scheme change still
transforms the coupling vector\foot{Capitalized indices run through all
couplings.  For matrices we use, e.g., $Q_I^{\phantom{I}\!J}$ for both
$Q_{ab}$ and $P_{ij}$.} $g^I$ and the wavefunction renormalization matrix
$Z_I^{\phantom{I}\!J}(g)$ as in \Schemechangeone but, due to the vector and
matrix character of the coupling and wavefunction renormalization
respectively, the new wavefunction renormalization is modified by a matrix
$F_I^{\phantom{I}\!J}(g)$ through
\eqn{Z^{1/2}(g)\to \widetilde{Z}^{1/2}(\tilde{g})=Z^{1/2}(g)F(g).}
Thus, under a scheme change, one has
\twoseqn{\tilde{\beta}^I(\tilde{g})&=\beta^J(g)\frac{\partial\tilde{g}^I}
{\partial g^J},}[RGnewoldA]{\tilde{\gamma}_I^{\phantom{I}\!J}(\tilde{g})&=
\left[F^{-1}(g)\gamma(g)F(g)\right]_I^{\phantom{I}\!J}
+\left[F^{-1}(g)\beta^K(g)\frac{\partial F(g)}{\partial g^K}
\right]_I^{\phantom{I}\!J}.}[RGnewoldB][RGnewold]
It is easy to see that, in the multi-coupling case, properties \ref{prop1}
and \ref{prop5} are still scheme-independent. Property \ref{prop2} is of
course modified so that only $\tra\gamma$ and $\det\gamma$, and so the
eigenvalues of $\gamma$, are scheme-independent. Property \ref{prop3} is
also modified since
\eqn{
\frac{\partial\tilde{\beta}^J(\tilde{g})}{\partial\tilde{g}^I}=
\frac{\partial g^K}{\partial\tilde{g}^I}\frac{\partial\beta^L(g)}{\partial
g^K}\frac{\partial\tilde{g}^J}{\partial g^L}+
\frac{\partial g^K}{\partial \tilde{g}^I}\beta^L(g)\frac{\partial}{\partial
g^L}\left(\frac{\partial\tilde{g}^J}{\partial g^K}\right),}[betader]
such that at a conformal fixed point the eigenvalues of
$\partial\beta^J(g)/\partial g^I$ are independent of the scheme.  This is
expected because $\partial \beta^J/\partial
g^I=\gamma_I^{\phantom{I}\!J}$, where $\gamma_I^{\phantom{I}\!J}$ is the
anomalous-dimension matrix of the operators sourced by the appropriate
couplings. Therefore, Eq.~\betader can be seen as an extension of
Eq.~\RGnewoldB with $F=\partial\tilde{g}/\partial g$.

Finally, if the one-loop beta function for one coupling depends on other
couplings, property \ref{prop4} is no longer true
\cite{Fortin:2011sz}---only the first coefficient in the beta function is
scheme-independent, although the UV or IR asymptotics of the couplings are
the same in any scheme.

\subsec{Natural scheme changes and scale-invariant trajectories}[NatScCh]
It is interesting to see how scale-invariant solutions behave under scheme
changes.\foot{The discussion of this subsection applies to scheme changes
under which Eqs.~\Scale transform covariantly.  Since the analysis for
gauge fields is straightforward, gauge fields are omitted for simplicity.}
Here we will distinguish between two types of scheme changes, which we dub
natural and unnatural.  A natural scheme change transforms the couplings as
\eqna{\lambda_{abcd}&\to\tilde{\lambda}_{abcd}=\lambda_{abcd}+\eta_{abcd},\\
y_{a|ij}&\to \tilde{y}_{a|ij}=y_{a|ij}+\xi_{a|ij},\\
y^\ast_{a|ij}&\to \tilde{y}^\ast_{a|ij}=
  y^\ast_{a|ij}+\xi^\ast_{a|ij},}[Naturalscheme]
such that all couplings transform covariantly with respect to the symmetry
group of the kinetic terms. MS and variants are examples of this---it
occurs, e.g., every time one dresses a Feynman diagram topology with
couplings.  Unnatural scheme changes spoil the covariance of equations.

We can now show that entries of $Q$ and $P$, which determine, e.g., the
frequency on a cyclic trajectory, are scheme-independent for natural scheme
changes.  Indeed, if the scheme change is natural, then the time evolution
of $\eta$ and $\xi$ on a scale-invariant trajectory is given by
\eqna{\eta_{abcd}(t)&=(e^{Qt})_{a'a}(e^{Qt})_{b'b}(e^{Qt})_{c'c}
(e^{Qt})_{d'd}\,\eta_{a'b'c'd'}(0),\\
\xi_{a|ij}(t)&=(e^{Qt})_{a'a}(e^{Pt})_{i'i}(e^{Pt})_{j'j}\,
\xi_{a'|i'j'}(0),}[SchemeShift]
and so
\eqn{\frac{d\eta_{abcd}}{dt}=Q_{a'a}\eta_{a'bcd}+\text{permutations},\qquad
\frac{d\xi_{a|ij}}{dt}=
  Q_{a'a}\xi_{a'|ij}+P_{i'i}\xi_{a|i'j}+P_{j'j}\xi_{a|ij'}.}[ShiftDer]
On a scale-invariant trajectory Eqs.~\Naturalscheme give
\eqn{\tilde{\beta}_{abcd}=\CQ_{abcd}-\frac{d\eta_{abcd}}{dt},\qquad
\tilde{\beta}_{a|ij}=\CP_{a|ij}-\frac{d\xi_{a|ij}}{dt},}[ScInvScheme]
and we can use Eqs.~\ShiftDer to obtain
\eqna{\tilde{\beta}_{abcd}&=
  -Q_{a'a}\tilde{\lambda}_{a'bcd}+\text{permutations},\\
\tilde{\beta}_{a|ij}&=
  -Q_{a'a}\tilde{y}_{a'|ij}-P_{i'i}\tilde{y}_{a|i'j}
  -P_{j'j}\tilde{y}_{a|ij'}.}
Hence, $Q$ and $P$ are scheme-independent for natural scheme changes.

As a result of our analysis the existence of scale-invariant trajectories
does not depend on the renormalization scheme. As expected, then, property
\ref{prop1} is easily extended to include non-conformal scale-invariant
trajectories.

Focusing on scalar anomalous dimensions (the argument can be easily
repeated for fermion anomalous dimensions), property \ref{prop2} can also
be generalized to scale-invariant theories.  Indeed, for natural scheme
changes on a scale-invariant trajectory Eq.~\RGnewoldB becomes
\eqn{\tilde{\gamma}_{ab}(\tilde{g})=\left[F^{-1}(g)\gamma(g)F(g)\right]_{ab}
+\left\{F^{-1}(g)[Q,F(g)]\right\}_{ab}}
since $-dF(g)/dt=[Q,F(g)]$.  One can then immediately see that (using
matrix notation)
\eqn{\tilde{\gamma}(\tilde{g})+Q=F^{-1}(g)[\gamma(g)+Q]F(g),}
so that the eigenvalues of $\gamma+Q$ are scheme-independent.  This is in
accord with expectations: in Ref.~\cite{Fortin:2011bm} it was shown that
the behavior of two-point functions is determined by the eigenvalues of
$\gamma+Q$, which are therefore expected to be scheme-independent.

Since property \ref{prop2} can be generalized to scale-invariant theories,
the same is expected for property \ref{prop3} due to $\partial
\beta^J/\partial g^I=\gamma_I^{\phantom{I}\!J}$. Indeed, Eq.~\betader
becomes
\eqn{\frac{\partial\tilde{\beta}^J}{\partial
\tilde{g}^I}=\left[F^{-1}(g)\frac{\partial\beta}
{\partial g}F(g)\right]_I^{\phantom{I}\!J}
+\left\{F^{-1}(g)[Q,F(g)]\right\}_I^{\phantom{I}\!J},}
where $F=\partial\tilde{g}/\partial g$, which gives (again using matrix
notation)
\eqn{\frac{\partial\tilde{\beta}}{\partial\tilde{g}}+Q=
F^{-1}(g)\left[\frac{\partial\beta}{\partial g}+Q\right]F(g).}
Therefore, the eigenvalues of $\partial\beta/\partial
g+Q=\partial(\beta-\CQ)/\partial g$ (since $\CQ=-gQ$) are
scheme-independent.  It is interesting to note that the eigenvalues of
$\partial\beta/\partial g+Q$ are expected to determine the character
(attractive, repulsive, etc.)\ of scale-invariant trajectories, and so one
of them should be zero---that is indeed the eigenvalue corresponding to the
(left) eigenvector $\beta^I=\CQ^I$.  This is because $\beta^I=\CQ^I$
generates a motion \emph{along} the scale-invariant trajectory, not away
from it, as can be seen directly from
\eqn{\beta^I\left[\frac{\partial\beta^J}{\partial g^I}+Q_I^{\phantom{I}\!J}
\right]_{\beta^I=\CQ^I}=-\frac{d\CQ^I}{dt}+\CQ^IQ_I^{\phantom{I}\!J}=0.}

Finally, properties \ref{prop4} and \ref{prop5} in the multi-coupling case
are trivially extended to scale-invariant theories since they do not rely
on the existence of scale-invariant trajectories (or conformal fixed
points).

To summarize, the scheme-independent properties (I--V) can be generalized
to:
\begin{enumerate}[label=(\Roman{enumi}$'$)]
    \renewcommand{\labelenumi}{(\Roman{enumi}$'$)}
  \item\label{prop1p} The existence of conformal fixed points and
  scale-invariant trajectories;
  \item\label{prop2p} The eigenvalues of $\gamma+Q$ at conformal fixed
  points and scale-invariant trajectories;
  \item\label{prop3p} The eigenvalues of $\partial\beta/\partial g+Q$ at
  conformal fixed points and scale-invariant trajectories;
  \item\label{prop4p} The first coefficient in the beta functions;
  \item\label{prop5p} The first coefficient in the anomalous-dimension
  matrix.
\end{enumerate}

\newsec{Stability properties}[stability]

\subsec{General discussion}
It is of interest to study the stability of scale-invariant solutions under
small deformations.  Such an analysis determines the character of a
particular scale-invariant solution, which can have (IR) attractive and/or
repulsive deformations.  In this section we will describe the properties of
all possible scale-invariant solutions.  The corresponding results for
conformal fixed points are recovered by setting $Q=0$ in the equations
below.  To simplify the equations, matrix notation is used throughout this
section.

Since non-conformal scale-invariant solutions exhibit non-trivial RG flows,
it is natural to disentangle the two contributions to the flow of the
deformations, i.e., the expected contribution from the non-conformal
scale-invariant solution, and the actual contribution from the deformations
which we want to analyze.  The appropriate quantity to study is thus
$\delta g(t)=[g(t)-g_*(t)]e^{-Qt}$, where $g_*(t)=g_*(0)e^{Qt}$ is a
scale-invariant solution, $\left.\beta\right|_{g=g_*(t)}=\CQ(t)$.  The
quantity $\delta g(t)$ determines the behavior of the deformations as a
function of RG time in a ``comoving frame'', i.e., \emph{modulo} the
expected non-conformal scale-invariant solution RG flow.  Note that,
although for non-conformal scale-invariant solutions the choice of $g_*(0)$
in $\delta g(t)=g(t)e^{-Qt}-g_*(0)$ is arbitrary,\foot{Any two points on a
non-conformal scale-invariant trajectory are physically equivalent due to
scale invariance.} in order to study the behavior of small deformations one
should first fix a $g_*(0)$.

To proceed further it is necessary to Taylor expand the beta functions
around the appropriate scale-invariant solution $g_*(t)$:
\eqn{\beta(t)=\left.\beta\right|_{g=g_*(t)}+[g(t)-g_*(t)]
\left.\frac{\partial\beta}{\partial g}\right|_{g=g_*(t)}+\cdots=
\CQ(t)+\delta g(t)\left.\frac{\partial\beta}{\partial g}
\right|_{g=g_*(0)}e^{Qt}+\cdots,}
where the last equality follows since $-d(\partial\beta/\partial g)/dt
=[Q,\partial\beta/\partial g]$ on the scale-invariant solution.  Note that
in order to disentangle the two contributions to the flow, the above Taylor
expansion is RG-time dependent.  It is now straightforward to write down,
at lowest non-trivial order, the system of (linear) differential equations
that the deformations must satisfy:
\eqn{-\frac{d\,\delta g(t)}{dt}=[\beta(t)-\CQ(t)]e^{-Qt}+\delta g(t)Q=
\delta g(t)S+\cdots,}[diffdeform]
where
\eqn{S=\left(\left.\frac{\partial\beta}
{\partial g}\right|_{g=g_*(0)}+Q\right)}
is the stability matrix. It is obvious that $\delta g(t)$ is the
appropriate choice of variable that allows a separation of the RG flow
contributions, for all RG-time dependence in Eq.~\diffdeform comes solely
from $\delta g(t)$.  Note, moreover, that Eq.~\diffdeform implies that the
behavior of the deformations $\delta g(t)$ is dictated by the eigenvalues
of $S$ which, as we showed in the previous section, are scheme-independent
(property \ref{prop3p}).   The solution to the system of differential
equations \diffdeform is simply
\eqn{\delta g(t)=\delta g(0)e^{-St}+\cdots}[deform]
and one can easily see that positive (respectively, negative) eigenvalues
of the stability matrix $S$ correspond to IR attractive (respectively,
repulsive) deformations.  As usual, the fate of deformations related to
vanishing eigenvalues cannot be determined from Eq.~\deform---for vanishing
eigenvalues it is necessary to go to higher order in the Taylor
expansion~\diffdeform.  However, as already mentioned, non-conformal
scale-invariant solutions exhibit one special (left) eigenvector $\delta
g(0)\propto\CQ(0)$ with vanishing eigenvalue which represents a deformation
along the scale-invariant solution.  For this special deformation the full
solution $\delta g(t)=[g_*(t\pm\delta t)-g_*(t)]e^{-Qt}=g_*(0)[e^{\pm
Q\,\delta t}-1]=\mp\CQ(0)\,\delta t+\cdots$ is RG-time independent as
expected, since it corresponds to a flow along the RG scale-invariant
trajectory.

The previous analysis is a generalization of the similar analysis done for
conformal solutions where $Q=0$.  Note that the special (left) eigenvector
$\delta g(0)\propto\CQ(0)$ does not exist for conformal fixed points, as
expected since conformal solutions do not exhibit any non-trivial RG flow.

\subsec{The example}
We can now use the results discussed above to investigate the behavior of
small deformations away from scale-invariant solutions.  To this end it is
natural to use an $\epsilon$ expansion for the stability matrix $S$ and its
eigenvalues $x_m$,
\eqn{S=\sum_{n\geq2}S^{(\frac{n}{2})}\epsilon^{\frac{n}{2}},\qquad
x_m=\sum_{n\geq2}x_m^{(\frac{n}{2})}\epsilon^{\frac{n}{2}}.}
The form of the expansion is dictated by the form of the beta functions in
the stability matrix.

The eigenvalues of the stability matrix are the roots of the characteristic
polynomial $\det(x\mathds{1}-S)$ which can also be expanded in $\epsilon$.
To lowest order the characteristic polynomial simplifies and the
eigenvalues are solutions of
\eqn{\det(x^{(1)}\mathds{1}-S^{(1)})=0.}[charloworder]
Since there are only seven non-vanishing independent couplings
($\lambda_{1,\ldots,5},y_{1,2}$ in \eqref{ttpotential}) at a generic point
on the non-conformal scale-invariant solution described in
section~\expansion,
Eq.~\charloworder for the corresponding couplings is
\begin{multline*}
z(z-1)\left(z^5-\tfrac{\sqrt{\num{52953}}}{57}z^4+\tfrac{\num{1894}
+\sqrt{\num{52953}}}{\num{475}}z^3
-\tfrac{\num{240768}-\num{335}\sqrt{\num{52953}}}{\num{135375}}z^2\right.\\
\left.-\tfrac{\num{421203}-\num{1573}\sqrt{\num{52953}}}{\num{225625}}z
+\tfrac{\num{136}(\num{757}\sqrt{\num{52953}}-\num{158859})}
{\num{64303125}}\right)=0
\end{multline*}
which cannot be solved by factorization into radicals.  (To avoid clutter
we define $z=x^{(1)}$.)  A numerical solution gives five positive, one
negative and one vanishing eigenvalue:
\eqn{z\approx2.4,\quad z=1,\quad z\approx0.99,\quad z\approx0.74,\quad
z\approx0.095,\quad z\approx-0.19,\quad z=0.}
The positive eigenvalues show that the scale-invariant solution is IR
attractive in several directions.  We thus expect that the limit cycle can
be reached by an appropriate deformation of a theory defined at a UV
conformal fixed point, although, to be certain, a more thorough analysis is
necessary.

\newsec{On the proof of the \texorpdfstring{$\mathbf{\emph{c}}$}{c}-theorem
at weak coupling}[c-thm]
As discussed in the introduction, our three-loop results do not contradict
the work of Osborn \cite{Osborn:1989td,Osborn:1991gm} and Jack \& Osborn
\cite{Jack:1990eb}.  Focusing on Ref.~\cite{Osborn:1991gm}, Osborn proved
that RG flows are gradient flows at two loops in the weak coupling regime.
Lifting the theory to curved space with spacetime-dependent couplings,
Osborn showed that Weyl consistency conditions lead to
\eqn{\frac{dc}{dt}=-\beta^I\frac{\partial c}{\partial
g^I}=-G_{IJ}\beta^I\beta^J,}[gradient]
with $G_{IJ}$ positive-definite in the weak coupling regime, thus
forbidding the existence of recurrent behaviors at all loops.  From the
analysis of Ref.~\cite{Osborn:1991gm} it would thus seem that
scale-invariant trajectories are forbidden to all orders in perturbation
theory.  However, the analysis of Ref.~\cite{Osborn:1991gm} leading to
Eq.~\gradient is too restrictive---it does not allow for spin-one operators
of dimension three, i.e., it does not include the possibility of
non-conformal scale-invariant theories.

The more general analysis, also performed by Osborn in
Ref.~\cite{Osborn:1991gm}, includes possible spin-one operators of
dimension three, which are related to the symmetry group of the kinetic
terms.  Such an analysis is done by promoting the related symmetry of the
kinetic terms---for example the symmetry of the kinetic terms generated by
the virial current, the natural spin-one operator of dimension three for
scale-invariant theories---to a symmetry of the interacting theory.  This
is implemented by allowing the couplings to transform appropriately under a
change generated by the spin-one operators of dimension three and by
introducing background gauge fields to render the symmetry local.  Then,
assuming that the regularization procedure preserves local gauge
invariance, Osborn's Weyl consistency conditions and current conservation
show that
\eqn{\frac{dc}{dt}=-\beta^I\frac{\partial c}{\partial
g^I}=-(G_{IJ}+\cdots)\beta^IB^J,}[ngradient]
where $B^I=\beta^I-\mathcal{Q}^I$.  Note that $B^I=0$ is precisely the
condition for scale invariance.  Thus, by allowing non-conformal
scale-invariant theories from the start, the work of
Refs.~\cite{Osborn:1989td,Osborn:1991gm,Jack:1990eb} implies the existence
of a $c$-function whose RG-time derivative vanishes at conformal fixed
points as well as on scale-invariant trajectories.  Note, moreover, that
the $c$-function might not be monotonically decreasing due to the extra
contributions to $dc/dt$ represented by the ellipsis in Eq.~\ngradient.

Note that, by promoting the symmetry of the spin-one operators of dimension
three to a symmetry of the interacting theory, it is natural to demand
regularization and renormalization schemes that satisfy the newly promoted
symmetry. This also explains the special status of the natural
renormalization schemes defined in the previous section.

Finally, it is interesting to see why the interference between quartic
coupling one-loop beta functions and Yukawa coupling two-loop beta
functions proposed by Wallace \& Zia \cite{Wallace:1974dy} as a possible
obstruction to the gradient flow interpretation of the RG flow is
circumvented by the introduction of the metric.  Focusing on the
problematic monomials in a possible $c$-function,
\eqn{c\supset
d_1\tra(y_a^*y_b^{\phantom{*}}y_c^*y_d^{\phantom{*}})\lambda_{abcd}+
d_2\tra(y_a^*y_b^{\phantom{*}})\lambda_{acde}\lambda_{bcde},}
the related contributions to the beta functions at one and two loops
respectively are
\eqna{\frac{\partial c}{\partial\lambda_{abcd}}&\supset
d_1\tra(y_a^*y_b^{\phantom{*}}y_c^*y_d^{\phantom{*}})+
2d_2\tra(y_d^*y_e^{\phantom{*}})\lambda_{abce}+\text{permutations},\\
\frac{\partial c}{\partial y_{a}}&\supset
2d_1y_b^{\phantom{*}}y_c^*y_d^{\phantom{*}}
  \lambda_{abcd}+d_2y_b\lambda_{acde}\lambda_{bcde}.}
Comparing with the true beta functions,
\eqna{\beta_{abcd}^{\text{(1-loop)}}&\supset-\frac{1}{16\pi^2}
\tra(y_a^*y_b^{\phantom{*}}y_c^*y_d^{\phantom{*}})
+\frac{1}{16\pi^2}\frac16\tra(y_d^*y_e^{\phantom{*}})
\lambda_{abce}+\text{permutations},\\
\beta_a^{\text{(2-loop)}}&\supset-\frac{2}{(16\pi^2)^2}\,
y_b^{\phantom{*}}y_c^*y_d^{\phantom{*}}\lambda_{abcd}
+\frac{1}{(16\pi^2)^2}\frac{1}{12}\,y_b\lambda_{acde}\lambda_{bcde},}
it is straightforward to see that the metric can account for the loop
mismatch since $d_2/d_1=-1/12$ for \emph{both} beta functions, as pointed
out in Ref.~\cite{Jack:1990eb}.  Note that the conditions for a gradient
flow interpretation of the RG flow introduced at higher orders are ever
more constraining due to the large number of diagrams\foot{This was already
noticed in Ref.~\cite{Dolan:1993vf}.} and it is plausible that they are not
satisfied, as our three-loop computation shows.  The interference argument
of Wallace \& Zia \cite{Wallace:1974dy} prevails at three loops, although
for a complete investigation the knowledge of the full three-loop beta
functions is necessary.  Interestingly, the interference between the
$(n-1)$-loop quartic-coupling beta function and the $n$-loop Yukawa beta
function also explains why the $n$-loop quartic-coupling beta function is
not necessary to argue for the existence of scale-invariant theories at
$n$-th order in perturbation theory.

\newsec{Cyclic trajectories and the
\texorpdfstring{$\mathbf{\emph{c}}$}{c}-theorem}[trajectories]
It is important to note that the existence of recurrent behaviors in RG
flows in $d=4$ does not contradict all versions of the
$c$-theorem.\foot{For a more extensive discussion see
Ref.~\cite{Fortin:2011sz}.}  In particular, the weak version of the
$c$-theorem, where two conformal fixed points connected by an RG flow
satisfy the inequality
\eqn{a_{\text{UV}}-a_{\text{IR}}>0}[atheorem]
with $a$ the conformal anomaly (see, for example,
Ref.~\cite{Cardy:1988cwa}),\foot{A claim for the proof of the inequality
\atheorem appeared recently in Ref.~\cite{Komargodski:2011vj} (see also
Ref.~\cite{Komargodski:2011xv}).} is consistent with scale without
conformal invariance.  Even the stronger version of the $c$-theorem, where
there exists a local function which is monotonically decreasing along
non-trivial RG flows, is compatible with recurrent behaviors \emph{as long
as the $c$-function is constant on scale-invariant trajectories}.  Only the
strongest version of the $c$-theorem is violated by the existence of limit
cycles and ergodicity; a gradient flow interpretation of RG flows is
impossible for theories in which scale does not imply conformal invariance.

Since theories exhibiting limit cycles or ergodicity are scale-invariant,
it is reasonable to expect the interpolating $c$-function to be constant on
scale-invariant trajectories.  Any such interpolating function is invariant
under the symmetry group of the kinetic terms, i.e., it does not carry
scalar or fermion indices.  Thus, in a natural scheme, all the explicit
RG-time dependence disappears on a scale-invariant trajectory.  This is the
behavior that is intuitively expected of the $c$-function, which should be
some measure of the number of massless degrees of freedom of the theory.
Therefore it must be constant on scale-invariant trajectories since any two
points on such trajectories are physically equivalent.

This behavior is very different from that encountered on cyclic flows
described in Ref.~\cite{LeClair:2003hj} and recently discussed in
association with the $c$-theorem in Ref.~\cite{Curtright:2011qg} (see also
Ref.~\cite{Morozov:2003ik}).  In Ref.~\cite{Curtright:2011qg}, the authors
argue that monotonic RG flows can be simultaneously cyclic if one allows
for a multi-valued interpolating $c$-function.  This is fundamentally
different from recurrent behavior with continuous scale invariance.  As
mentioned above, the interpolating $c$-function must be constant on
scale-invariant trajectories.  Moreover, the examples cited in
Ref.~\cite{Curtright:2011qg} exhibit one feature, turning points, which
does not appear on continuously scale-invariant trajectories.  Turning
points are peculiar locations in coupling space: the beta functions vanish
there, but the first derivative of the beta functions diverges.
Consequently, RG flows can overshoot turning points.  In contrast, all
existing continuously scale-invariant examples are well-defined smooth
weakly-coupled theories, and thus do not display turning points.  The
existence of turning points on cyclic flows is a reflection of the
possibility of multi-valued $c$-functions which are monotonically
decreasing along the flow.  Here we want to stress that the physics of
cyclic flows with turning points as described in
Ref.~\cite{Curtright:2011qg} is very different from that of recurrent
behaviors with continuous scale invariance.  It is therefore very unlikely
that monotonically decreasing multi-valued $c$-functions exist on
scale-invariant recurrent behaviors as suggested in
Ref.~\cite{Curtright:2011qg}.

\newsec{Conclusion}[conclusion]
Does scale imply conformal invariance in unitary relativistic QFTs?  The
answer is negative in $d=4-\epsilon$.  Although a similarly conclusive
statement in the $d=4$ case cannot yet be made, we strongly believe that
the answer there is also negative.  There are no physical arguments on
which one can rely to forbid non-conformal scale-invariant theories.
Instead, one simply needs to compute the beta functions and explore the
different regions in coupling space.  That an example of a scale-invariant
theory which is not conformal eluded the physics community for so long is
easily explained by the complexity of the problem: to see non-conformal
scale-invariant theories, one must go to three loops, and the beta
functions at three loops in the most general QFT are not known.

\ack{We would like to thank Aneesh Manohar for helpful discussions and Ken
Intriligator for useful discussions and comments on the manuscript.  We
also thank Miko{\l}aj Misiak for guiding us through the literature on
three-loop calculations.  Finally, we are especially grateful to Hugh
Osborn for enlightening correspondence and for his comments on the
manuscript. This work was supported in part by the US Department of Energy
under contract DOE-FG03-97ER40546.}

\bibliography{3LoopSvC_ref}

\end{document}